\documentclass[12pt]{article}

\usepackage{graphicx}
\usepackage{epstopdf}

\newcommand{\beq}{\begin{equation}}
\newcommand{\eeq}{\end{equation}}
\newcommand{\beqa}{\begin{eqnarray}}
\newcommand{\eeqa}{\end{eqnarray}}
\newcommand{\beqar}{\begin{eqnarray*}}
\newcommand{\eeqar}{\end{eqnarray*}}

\begin{document}
\thispagestyle{empty}

\hfill{\sc UG-FT-247/09}

\vspace*{-2mm}
\hfill{\sc CAFPE-117/09}

\vspace{32pt}
\begin{center}

\textbf{\Large Cosmic ray knee and new physics at the TeV scale\\
} 
\vspace{40pt}

Roberto Barcel\'o$^{1}$, Manuel Masip$^{1}$, 
Iacopo Mastromatteo$^{2}$
\vspace{12pt}

\textit{
$^{1}$CAFPE and Departamento de F{\'\i}sica Te\'orica y del
Cosmos}\\ 
\textit{Universidad de Granada, E-18071 Granada, Spain}\\
\vspace{8pt}
\textit{$^{2}$International School for Advanced Studies (SISSA)}\\ 
\textit{Via Beirut 2-4, I-34014 Trieste, Italy}\\
\vspace{16pt}
\texttt{rbarcelo@ugr.es, masip@ugr.es, mastroma@sissa.it}
\end{center}

\vspace{40pt}

\date{\today}

\begin{abstract}

We analyze the possibility that the cosmic ray knee appears
at an energy threshold where the proton--dark matter 
cross section becomes large due to new TeV physics. 
It has been shown that 
such interactions could {\it break} the proton and produce
a diffuse gamma ray flux consistent with MILAGRO observations.
We argue that this hypothesis implies {\it knees} that scale 
with the atomic mass for the different nuclei, as
KASKADE data seem to indicate.
We find that to explain the change in the 
spectral index in the flux from $E^{-2.7}$
to $E^{-3.1}$ the cross section must grow like 
$E^{0.4+\beta}$ above the knee, where $\beta=0.3$--0.6 parametrizes
the energy dependence of the {\it age}  
($\tau \propto E^{-\beta}$) of the cosmic 
rays reaching the Earth. The hypothesis 
also requires mbarn cross sections (that could be modelled 
with TeV gravity) and large densities of 
dark matter (that could be clumped around the 
sources of cosmic rays). 
We argue that neutrinos would also exhibit a
threshold at $E= (m_\chi/m_p)\; E_{knee}\approx 
10^8$ GeV where their interaction with a nucleon becomes
strong. Therefore, the observation at ICECUBE or ANITA 
of standard neutrino events 
above this threshold would disprove the scenario.

\end{abstract}

\newpage

\section{Introduction}

The observed cosmic ray flux reaching the Earth extends 
up to energies around $10^{11}$ GeV. 
It seems remarkable that 
over 10 decades of energy this flux can be described in 
very simple terms \cite{Amsler:2008zzb}. Between 10 and $10^6$ GeV 
(the {\it knee}) it is given by 
\beq
\frac{{\rm d}\Phi_{N}}{{\rm d}E}\approx 
1.8 \; E^{-2.7}\; {\rm {nucleons\over cm^2\;s\;sr\;GeV}}\;.
\label{pcr}
\eeq
At $E_{knee}$ the spectral index changes to $-3.1$ and stays
constant up to $\approx 10^{9.5}$ GeV (the {\it ankle}). There 
it goes back to $-2.7$ until (arguably) hits the GZK
suppression a decade later 
(see \cite{Hillas:2006ms,Dar:2006dy} for a review). 

The ankle is generally explained as the overtaking in the 
cosmic ray flux of a new component of extragalactic origin. 
These ultra-high energy particles
arrive from all regions in the sky, which excludes the possibility
that they are produced only in the galactic disc.
For less energetic particles, $E\le 10^8$ GeV, one can assume 
production in the disc despite
the fact that the observed flux is also isotropic. The reason is 
that their Larmor radius is smaller, so the
trajectory is isotropized by the random magnetic fields present
in our galaxy ($r_L\approx 0.1$ kpc for a $E=10^8$ GeV proton 
inside a $B=1\;\mu$G magnetic field).

The other main feature in the spectrum, the knee, seems more
involved. It is usually thought that $E_{knee}$ reflects the maximum 
energy reached by the dominant cosmic accelerators 
in our galaxy. In that case, however, one would 
expect a transition regime (not seen) between $E_{knee}$
and the energy $E$ where the new acceleration mechanism
dominates. Propagation and confinement within
the galaxy could also play an important role. In particular, 
the knee could correspond to a critical energy where a large
fraction of cosmic rays escapes the galactic disc before
reaching the Earth \cite{Prouza:2003yf}. Again, it seems 
difficult to obtain a
sustained spectrum $\propto E^{-3.1}$ from that energy up to 
$E_{ankle}$ \cite{Aloisio:2006wv}. 

In this paper we explore a different possibility recently 
proposed 
in \cite{Masip:2008mk}. Namely, the knee could be caused
by new interactions with dark matter particles ($\chi$)
in our galaxy. This hypothesis requires that the 
cross section $\sigma_{p\chi}$ 
becomes strong and {\it breaks} the cosmic 
protons of energy above the threshold $E_{knee}$
in their way to the Earth. It is 
supported by the possible excess in the 
diffuse gamma-ray flux observed by MILAGRO \cite{Abdo:2008if}.
Here we study other possible implications and also the
conditions (on the cross section and the dark matter 
distribution) that have to be satisfied for this scenario 
to really work.

\section{Other knees and the change in the spectral index}

In addition to protons, the cosmic ray flux 
contains atomic nuclei (He, C, Fe, etc.). 
The abundance of
the different species at energies around $10^6$ GeV has been
measured in extensive air-shower experiments. 
In particular, KASCADE \cite{Antoni:2005wq,Aglietta:2004np} 
has observed that these 
fluxes also seem to exhibit a knee at an energy that increases
with the atomic number. 
It is obvious
that if there is an energy threshold $E_{knee}$
where the $p$--$\chi$ cross section becomes strong and produces
the proton knee, also a nucleus $N$
will experience a similar effect.

As a first approximation, if $\sigma_{p\chi}$
is negligible at energies 
below $E_{knee}$ then the nucleus will start interacting 
when the energy
per nucleon reaches that same threshold. This implies that
the knee scales
linearly with the atomic mass $A$:
\beq
E_{knee}^N\approx A\; E_{knee}
\;.
\eeq
One can take into account, however, that $\sigma_{p\chi}$
is not zero at $E<E_{knee}$ (a step function), and that 
its raise with the energy may be better 
described by the power law
\beq
\sigma_{p\chi}(E)\propto E^\alpha\;\;\;\; (E<E_{knee})
\;.
\eeq
In that case the atomic knee 
is moved towards the proton knee $E_{knee}$
by two factors. First, the electric charge $Z>1$ of a
nucleus increases its interaction strength 
with the turbulent magnetic
fields in our galaxy \cite{Han:2009ts}. 
This implies that the average time 
$\tau$ (and distance $L=c\tau$)
that it is travelling from the source to the Earth
also increases \cite{Strong:2007nh}:
\beq
\tau \propto R^{-\beta}
\;,
\label{time}
\eeq
where $R=E/Z$ is the rigidity and the index $\beta$ at
these energies may take values between 0.3 (typical for a 
Kolmogorov spectrum of magnetic fluctuations) and 0.6 
(as deduced from the study of fluxes of 
stable and unstable nuclei
of energy below the TeV). The probability that a cosmic
ray interacts along its trajectory grows with $\tau$,
\beq
p_{int} \approx 1 - 
e^{-\sigma_{N\chi} n_\chi L}\approx 
\sigma_{N\chi}\;n_\chi\; c\tau
\;,
\label{prob}
\eeq
where $n_\chi=\rho_\chi/m_\chi$ is the number density
of dark matter particles (the average depth $x$ of a 
trajectory is $\rho L$, and the nucleus interaction length 
is $1/(\sigma_{N\chi}n_\chi)$). Therefore, a larger $\tau$
requires a smaller $\sigma_{N\chi}$ to obtain the
same probability of interaction.
The second factor is that the nucleus-dark matter cross
section is larger than $\sigma_{p\chi}$. One can estimate
that 
\beq
\sigma_{N\chi} (E) \approx \left\{
\begin{array}{l l} 
\displaystyle A\;\sigma_{p\chi}(E/A)
& \sigma_{p\chi} \le 1\; {\rm mbarn}\;; \\
\displaystyle A^{2/3}\;\sigma_{p\chi}(E/A)
& \sigma_{p\chi} \gg 1\; {\rm mbarn}
\;.
\end{array} \right. 
\eeq
Therefore, the interaction probability at the nucleus knee 
will coincide with the one at the proton knee if 
\beq
E_{knee}^N \approx E_{knee}\; 
\displaystyle 
A^{\alpha-1\over \alpha-\beta} Z^{-\beta\over \alpha-\beta}
\;.
\label{atom}
\eeq
For large values of $\alpha$ we recover the linear scaling 
with the mass number, whereas 
taking $\alpha=2$ and $\beta=0.5$, the helium, 
carbon, and iron knees would move from 
$4 E_{knee}$, $12 E_{knee}$ and $55 E_{knee}$
to $2 E_{knee}$, $3 E_{knee}$and $5 E_{knee}$, 
respectively\footnote{Intermediate values in the position 
of the knee would be
obtained for larger values of $\alpha$}.
Therefore, we conclude that 
the hypothesis under study here seems able to explain the 
knees for the different nuclei observed by KASKADE.

Another important point to address is whether these
interactions could also explain the 
constant spectral index $-3.1$ in the flux 
between the knee and the ankle.
Let us suppose that with no dark matter perturbing its
propagation the cosmic ray
flux would have followed a $E^{-2.7}$ dependence up
to $E_{ankle}$. This implies that the probability of 
interaction changing the spectral index from $-2.7$ 
to $-3.1$ must be
\beq
p_{int}(E>E_{knee}) \approx 1 - \left( {E\over E_{knee}} 
\right)^{-0.4}
\;.
\eeq
We will show that 
this probability can be obtained if 
above $E_{knee}$ the proton-dark matter
cross section keeps growing with a given exponent $\alpha'$:
\beq
\sigma_{p\chi}\propto E^{\alpha'}\;\;\;\; (E>E_{knee})
\;.
\eeq
The probability of interaction for a cosmic ray along
a trajectory of length $l$ is
\beq
p_{int}(l, E) = 1 - e^{-\sigma_{p\chi}n_\chi l} 
\;.
\label{prob3}
\eeq
Now let us assume that $l$ is a random variable
with an exponential probability distribution $w(l)$ and
an average value $\langle l \rangle=L$:
\beq
w(l) = {1\over L} e^{-l/L} 
\;.
\eeq
This is justified if the cosmic ray propagation can
be modelled by a diffusive process driven by irregularities
in the galactic magnetic field \cite{Strong:2007nh}. 
The diffusion coeficient is
$D\sim L^{-1}\propto E^\beta$ and lengths larger than $L$ 
will be suppressed exponentially. The probability of 
interaction is then
\beq
p_{int}(E) =\int {\rm d}l\; p_{int}(l, E)\; w(l) =
1-{1\over \sigma_{p\chi} n_\chi L + 1}
\;. 
\label{prob4}
\eeq
Since at energies above $E_{knee}$ 
$\sigma_{p\chi} n_\chi L$ grows fast larger than 1, 
a constant spectral index needs that 
\beq
\left( {E\over E_{knee}} \right)^{-0.4}\approx 
\sigma_{p\chi} n_\chi L \propto
E^{\alpha'-\beta}
\;.
\label{prob5}
\eeq
If $\sigma_{p\chi}$ grows between the knee
and the ankle like $E^{0.4+\beta}$, with 
$\beta= 0.3$--0.6, then we can explain a sustained
spectral index of $-3.1$ in the cosmic ray flux.

\section{Constraints and predictions at neutrino telescopes}

The large probability of interaction required to produce the
knee means that
\beq
\sigma_{p\chi}n_\chi L =
{\sigma_{p\chi}x_\chi\over m_\chi} \approx 1
\;.
\label{prob6}
\eeq
It is easy to see that this can only be achieved with 
very large values of the cross section $\sigma_{p\chi}$ and 
of the dark matter density $\rho_\chi$. In particular, we know that
as cosmic rays propagate from the source a significant fraction
of them interact with interestellar matter. Since they collide
with a hadronic cross section, the amount of matter that they 
cross must be {\it a few} g/cm$^2$ (a hadronic interaction length 
is around 50 g/cm$^2$ \cite{Amsler:2008zzb}). 
Therefore, on dimensional grounds 
$\chi$ may have an impact on the cosmic ray propagation
if $\sigma_{p\chi}$ grows above 
the mbarn and $\rho_\chi$ is larger than the density of 
interstellar gas.

Let us first discuss the cross section. A possibility that has
been extensively discussed during the past years is strong TeV 
gravity in models with extra dimensions \cite{ArkaniHamed:1998rs}. 
If the fundamental scale is $M_D\approx 1$ TeV, at larger center of 
mass energies the gravitational cross section will dominate over
gauge boson exchange due to the spin 2 of the graviton. At 
lower energies (below the mass $M_c$ of the first Kaluza-Klein 
mode)
4-dimensional gravity is unchanged, whereas between $M_c$ and
$M_D$ Newton's constant grows like a power law. An analysis of
the proton--dark matter cross section at center if mass
energies $s>M_D^2$ 
in these models can be found in \cite{Illana:2005pu}. 
The scattering
is dominated by eikonal processes where a parton carrying a
fraction $x$ of the proton momentum transfers a
{\it small} fraction $y$ of energy to 
$\chi$. The cross section at the parton level 
grows\footnote{Notice that growth of  
$\sigma_{p\chi}$ with the energy will also be 
determined by the parton distribution functions.}
like $\hat s^{1+4/n}$, so it is larger and softer 
(involves larger distances and smaller energy transfer) 
for lower values of the number $n$ of extra dimensions.
This is easily understood because gravity
{\it dilutes} faster with the distance for larger $n$, 
if $n<2$ it
is a long distance interaction with a divergent total 
cross section. 

\begin{figure}
\begin{center}
\includegraphics[width=0.5\linewidth]{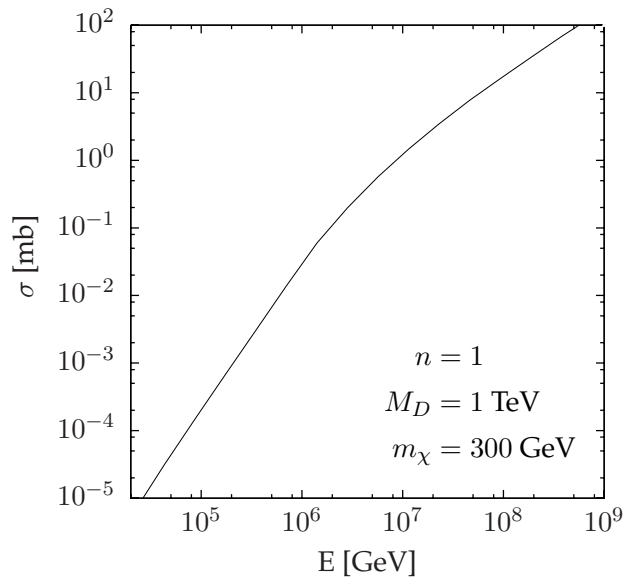}
\end{center}
\caption{Cross section for an eikonal gravitational  
interaction in $p$--$\chi$ 
collisions.
\label{fig1}}
\end{figure}
We find that if $M_D\ge 1$ TeV only $n=1$ seems to give 
the large values of $\sigma_{p\chi}$
required to explain the cosmic ray knee. Of course, we need
in that case a mechanism to avoid bounds from supernova
and from macroscopic gravity. For example, 
Giudice et al \cite{Giudice:2004mg}
build a model where a warp factor {\it adds} a mass of 
$\approx 10$ MeV to the Kaluza-Klein excitations of the
graviton without changing the basic properties of the
5-dimensional model at high energies. The effective
model has a TeV fundamental scale and only one extra dimension
up to distances of $\approx (10\;{\rm MeV})^{-1}$.
In Fig.~1 we plot the cross section in this setup. The
average fraction of energy transferred by a $E=10^6$ GeV 
proton to the dark matter particle in an interaction is
$\langle y \rangle=10^{-5}$. We have required a minimum
$q^2$ of 1 GeV$^2$, since the collision must break the 
incident proton.

Let us now comment on the dark matter density that is
required. Although it is thought that the total amount 
of dark matter in the galaxy is much larger than the 
amount of baryonic interstellar matter, the former would
be distributed inside a spherical halo of 
$\approx 200$ kpc \cite{Navarro:1995iw}, 
whereas the latter would be mainly within a 
disc of thickness $\approx 6$ kpc. As a consequence,
the depth $x_\chi$ of dark matter crossed by a cosmic ray from
the source (in the galactic disc) to the Earth would
be similar or even 
smaller than that of baryonic gas ($x_{IM}$). For example, if
we assume a constant  
$\rho_\chi \approx 0.3$ g/cm$^3$ (the local density
near the solar system) then $x_\chi\approx 0.1 x_{IM}$.
This value is clearly insufficient to explain the 
cosmic ray knee. 

We think, however, that one may consider scenarios where 
the density $\rho_\chi$ that cosmic rays face in their
way to the Earth is substantially larger 
($x_\chi\approx 100\; x_{IM}$). In particular, the dark matter 
could be more clumped locally, in the regions where the 
cosmic rays are produced. It is also possible that 
cosmic rays spend a significant fraction of
time inside local clouds of dark matter, trapped by stronger 
magnetic fields. They would also face a larger depth 
if the distribution of dark matter were more flattened  
towards the galactic plane (a non-sperical halo)
\cite{Sackett:1994va}. This third 
possibility might be better accommodated if the galactic 
dark matter has two components (a {\it heavier} and
a {\it lighter} one \cite{Hsieh:2006fg}), as the proximity 
of the heavier component to the fundamental scale in these 
models would imply a larger $\chi$--$\chi$ elastic cross 
section justifying an anomalous distribution. Notice that 
although the total amount of dark matter in the galaxy is
well established, its distribution is {\it underconstrained} 
(the rotation curves can be fitted with 
different distributions \cite{Battaner:2000ef,Martins:2009cp}, 
and magnetic fields could also play a role {\cite{ednature}).

\begin{figure}
\begin{center}
\includegraphics[width=0.54\linewidth]{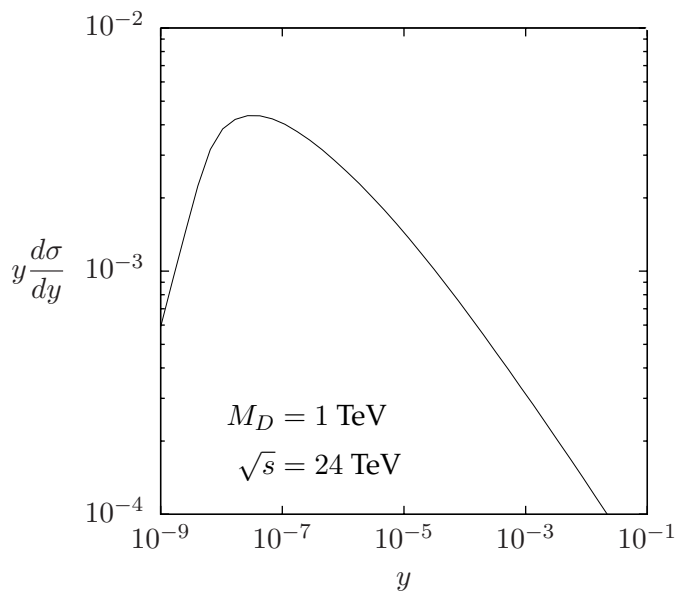}
\end{center}
\caption{Differential cross section $d\sigma/dy$ (mbarn)
in a $\nu$--$p$ (or $p$--$\chi$) gravitational 
interaction, where $y=(E_\nu-E'_\nu)/E_\nu$ is the
fraction of energy lost by the incident neutrino (or
transferred to $\chi$). 
\label{fig2}}
\end{figure}
We would like to discuss a last consequence 
that could be used to disregard the scenario under study
in a model independent way.
If gravity produces an increase in the dark matter
interactions, it will also imply a large cross section for
neutrinos above a certain energy threshold. More precisely,
the center of mass energy when a neutrino of energy $E_\nu$
hits a proton at rest is
\beq
\sqrt{s}=\sqrt{2 m_p E_\nu}
\;. 
\eeq
This coincides with the $\sqrt{s}$ in the collision
of a proton at the knee with a dark matter particle if
\beq
E_\nu^{th}={m_\chi\over m_p}\; 
E_{knee}\approx 10^8\;{\rm GeV}
\;. 
\eeq
Therefore, in this framework we would not observe 
any neutrino events above $E_\nu^{th}$ at $\nu$--telescopes. 
Such a neutrino 
would have an interaction lenght of 1--10 m in rock, 
losing a $y\approx 10^{-5}$ fraction of its energy in each collision.
In Fig.~2 we plot the distribution of $y$ in 
neutrino--proton interactions.

For example, a cosmogenic neutrino \cite{Semikoz:2003wv} 
of $10^9$ GeV would interact with a mbarn cross section and
lose around 1 TeV of energy ($y\approx 10^{-6}$) 
in each interaction with
matter. Energy loss, however, is dominated
by the (less frequent) collisions of larger $y$. We estimate
that such neutrino could deposit a $10\%$ of its energy
every 1--10 km.
This means that it would never reach a 
telescope from up-going directions. One would have, however,
other possible signals there. The neutrino could reach 
the center of IceCube (2 km under the ice) \cite{Achterberg:2006md}
from large (inclined) zenith angles. 
It would start there a TeV
hadronic shower every 1--10 meters of ice, an event that would be 
clearly different from the muon bundle produced by an extensive
air shower.

Notice also that bounds on the $\nu$--$p$ cross section from
the non-observation of quasi-horizontal extended air showers 
\cite{Anchordoqui:2004ma} produced by cosmogenic neutrinos 
do not apply here. The reason is that in the interaction the
neutrino will deposit just a small fraction $y$ of energy,  
starting an air shower below the 
$\approx 10^8$ GeV threshold in these experiments.

\section{Summary and discussion}

Cosmic rays may be affected by new physics at the TeV
scale. In particular, their interaction with dark matter
seems interesting because of two generic reasons:
{\it i)} the dark matter particle is expected to be
heavy, which provides larger center of mass energies 
($\sqrt{2m_\chi E}$) than in cosmic ray collisions with 
a nucleon, and  {\it ii)} the standard interactions  
are expected to be weak, which makes the relative
effect of new physics easier to detect.

We have studied the possibility that the knee 
observed in the cosmic ray flux at energies around
$10^6$ GeV may be caused by these interactions. In
particular, models with extra dimensions and a TeV
fundamental scale of gravity predict cross sections 
that grow very fast in the {\it trans-Planckian} regime. 
In \cite{Masip:2008mk} it is shown that these interactions 
could break the protons and produce a diffuse flux
of secondary gamma rays consistent with 
MILAGRO observations.
Here we have found that the knee  
would also imply knees that scale with the
atomic number for the other nuclei present in
the cosmic ray flux. We have shown that the
constant power law $E^{-3.1}$ for the flux 
between the knee and the ankle could
be correlated with a sustained growth 
$\sigma_{p\chi}\propto E^{\alpha'}$ in the 
proton--dark matter cross section.

We have also found, however, that the cross
sections and the dark matter densities which are 
required are very large (mbarn and 100 times 
over the expected value, respectively). 
In particular, the size
and the scaling of $\sigma_{p\chi}$ seem possible
only in models with one extra dimension. Such
a model 
should be completed with a mechanism increasing
the Kaluza-Klein masses like the one
discussed in \cite{Giudice:2004mg} to
be consistent with observations. As for the dark
matter, it may require a large local concentration
near the sources of cosmic rays, or a flattened 
galactic distribution. We have speculated
that this could be justified in a 2-component model 
(see, for example, \cite{Hsieh:2006fg}) where 
the heavier one may not be completely 
{\it collisionless} (the elastic scattering with
another dark matter particle could have a center of
mass energy just below the fundamental scale $M_D$).

Notice also that the presence of a possible 
{\it second knee} in the cosmic ray spectrum at 
$10^{8.5}$ GeV would not alter or invalidate 
the proposed mechanism.
This second knee would have a different origin, 
the dominance in the spectrum of extragalactic 
protons \cite{Berezinsky:2006mk,Aloisio:2007rc}. 
The extragalactic comic ray flux would not be affected by 
the interactions producing the first knee because 
dark matter densities 
are smaller outside the galaxy and the trajectories 
at these energies are {\it not} a random
walk \cite{Masip:2008mk}.

In any case, the scenario that we have discussed
has what we think is a 
model independent implication in neutrino physics
that could prove it wrong:
the absence of standard neutrino interactions 
above a threshold energy around $10^8$ GeV. 
Neutrinos above this energy would interact with
protons with a mbarn cross section, losing a 
small amount of energy (around 1 TeV) in each 
interaction. When a cosmogenic neutrino enters
the atmosphere horizontally, 
the penetrating shower that it starts would contain 
an energy below 
the triggers in extended air shower experiments. 
However, in $\nu$-telescopes one could
observe inclined events (down-going but 
from large zenith angles) 
where a very energetic neutrino starts a continuous
of showers spaced by 1--10 m,
a signal with no background within 
the standard model. The correlation between dark
matter and neutrino physics is a generic feature
in TeV gravity models. We find very appealing
that a possible observational effect, 
the knee from cosmic ray--dark matter
collisions, can be disproved (or supported) 
in the near future by another one, the 
observation of cosmogenic neutrino interactions at
ICECUBE or ANITA \cite{Gorham:2008yk}.

\section*{Acknowledgments}
We would like to thank Eduardo Battaner, Andrea 
Chiavassa, Mark Jenkins and Ute Lisenfeld for very useful discussions.
This work has been supported by MEC of Spain 
(FPA2006-05294) and by 
Junta de Andaluc\'\i a (FQM-101 and FQM-437).
RB acknowledges a FPI fellowship from MEC of Spain.
IM acknowledges a fellowship from SISSA (2007JHLPEZ).


\begin{thebibliography}{99}

\bibitem{Amsler:2008zzb}
  C.~Amsler {\it et al.}  [Particle Data Group],
  Phys.\ Lett.\  B {\bf 667} (2008) 1.

\bibitem{Hillas:2006ms}
  A.~M.~Hillas,
  arXiv:astro-ph/0607109.

\bibitem{Dar:2006dy}
  A.~Dar and A.~De Rujula,
  Phys.\ Rept.\  {\bf 466} (2008) 179.

\bibitem{Prouza:2003yf}
  M.~Prouza and R.~Smida,
  Astron.\ Astrophys.\  {\bf 410} (2003) 1.

\bibitem{Aloisio:2006wv}
  R.~Aloisio, V.~Berezinsky, P.~Blasi, A.~Gazizov, S.~Grigorieva and B.~Hnatyk,
  Astropart.\ Phys.\  {\bf 27} (2007) 76.

\bibitem{Masip:2008mk}
  M.~Masip and I.~Mastromatteo,
  JCAP {\bf 0812} (2008) 003.

\bibitem{Abdo:2008if}
  A.~A.~Abdo {\it et al.},
  ``A Measurement of the Spatial Distribution of Diffuse TeV Gamma Ray Emission
  from the Galactic Plane with Milagro,''
  arXiv:0805.0417 [astro-ph].

\bibitem{Antoni:2005wq}
  T.~Antoni {\it et al.}  [The KASCADE Collaboration],
  Astropart.\ Phys.\  {\bf 24} (2005) 1.

\bibitem{Aglietta:2004np}
  M.~Aglietta {\it et al.}  [EAS-TOP Collaboration],
  Astropart.\ Phys.\  {\bf 21} (2004) 583.

\bibitem{Han:2009ts}
  J.~L.~Han,
  arXiv:0901.1165 [astro-ph].

\bibitem{Strong:2007nh}
  A.~W.~Strong, I.~V.~Moskalenko and V.~S.~Ptuskin,
  Ann.\ Rev.\ Nucl.\ Part.\ Sci.\  {\bf 57} (2007) 285.

\bibitem{ArkaniHamed:1998rs}
  N.~Arkani-Hamed, S.~Dimopoulos and G.~R.~Dvali,
  Phys.\ Lett.\  B {\bf 429} (1998) 263;
  I.~Antoniadis, N.~Arkani-Hamed, S.~Dimopoulos and G.~R.~Dvali,
  Phys.\ Lett.\  B {\bf 436} (1998) 257;
  L.~Randall and R.~Sundrum,
  Phys.\ Rev.\ Lett.\  {\bf 83} (1999) 3370.

\bibitem{Illana:2005pu}
  G.~F.~Giudice, R.~Rattazzi and J.~D.~Wells,
  Nucl.\ Phys.\  B {\bf 630} (2002) 293;
  J.~I.~Illana, M.~Masip and D.~Meloni,
  Phys.\ Rev.\  D {\bf 72} (2005) 024003;
  E.~M.~Sessolo and D.~W.~McKay,
  Phys.\ Lett.\  B {\bf 668} (2008) 396;
  P.~Draggiotis, M.~Masip and I.~Mastromatteo,
  JCAP {\bf 07} (2008) 014.

\bibitem{Giudice:2004mg}
  G.~F.~Giudice, T.~Plehn and A.~Strumia,
  Nucl.\ Phys.\  B {\bf 706} (2005) 455.

\bibitem{Navarro:1995iw}
  J.~F.~Navarro, C.~S.~Frenk and S.~D.~M.~White,
  Astrophys.\ J.\  {\bf 462} (1996) 563.

\bibitem{Sackett:1994va}
  P.~D.~Sackett, H.~W.~Rix, B.~J.~Jarvis and K.~C.~Freeman,
  Astrophys.\ J.\  {\bf 436} (1994) 629.

\bibitem{Hsieh:2006fg}
  K.~Hsieh, R.~N.~Mohapatra and S.~Nasri,
  JHEP {\bf 0612} (2006) 067.

\bibitem{Battaner:2000ef}
  E.~Battaner and E.~Florido,
  Fund.\ Cosmic Phys.\  {\bf 21} (2000) 1.

\bibitem{Martins:2009cp}
  C.~F.~Martins,
  ``The distribution of the dark matter in galaxies as the imprint of its
  Nature,''
  arXiv:0903.4588 [astro-ph.CO].

\bibitem{ednature}
  E.~Battaner, J.L.~Garrido, M.~Membrado and E.~Florido,
  Nature {\bf 360} (1992) 652; 
  E.~Battaner and E.~Florido,
  Astron. Nachr. {\bf 328} (2007) 92.


\bibitem{Semikoz:2003wv}
  D.~V.~Semikoz and G.~Sigl,
  JCAP {\bf 0404} (2004) 003.

\bibitem{Achterberg:2006md}
  A.~Achterberg {\it et al.}  [IceCube Collaboration],
  Astropart.\ Phys.\  {\bf 26} (2006) 155.

\bibitem{Anchordoqui:2004ma}
  L.~A.~Anchordoqui, Z.~Fodor, S.~D.~Katz, A.~Ringwald and H.~Tu,
  JCAP {\bf 0506} (2005) 013.

\bibitem{Berezinsky:2006mk}
  V.~Berezinsky, A.~Gazizov and M.~Kachelriess,
  Phys.\ Rev.\ Lett.\  {\bf 97} (2006) 231101.

\bibitem{Aloisio:2007rc}
  R.~Aloisio, V.~Berezinsky, P.~Blasi and S.~Ostapchenko,
  Phys.\ Rev.\  D {\bf 77} (2008) 025007.

\bibitem{Gorham:2008yk}
  P.~W.~Gorham {\it et al.}  [ANITA collaboration],
  ``New Limits on the Ultra-high Energy Cosmic Neutrino Flux from the ANITA
  Experiment,''
  arXiv:0812.2715 [astro-ph].


\end{thebibliography}
\end{document}